# Origin of the exotic electronic states in antiferromagnetic NdSb

Peng Li[1†], Tongrui Li[2†], Sen Liao[1], Zhipeng Cao[3], Rui Xu[1], Yuzhe Wang[1], Jianghao Yao[1], Shengtao Cui[2], Zhe Sun[2], Yilin Wang[1], Xiangang Wan[3*], Juan Jiang[1*], Donglai Feng[1],

[1]*School of Emerging Technology and Department of Physics, University of Science and Technology of China, Hefei 230026, China*

[2]*National Synchrotron Radiation Laboratory, University of Science and Technology of China, Hefei 230026, People's Republic of China*

[3]*National Laboratory of Solid State Microstructures, School of Physics, Collaborative Innovation Center of Advanced Microstructures, Nanjing University, Nanjing 210093, China;*

[†]These authors contributed equally to this work.

E-mail addresses: xgwan@nju.edu.cn, jjiangcindy@ustc.edu.cn

**Abstract**

Using angle resolved photoemission spectroscopy measurements and first principle calculations, we report that the possible unconventional 2q antiferromagnetic (AFM) order in NdSb can induce unusual modulation on its electronic structure. The obvious extra bands observed in the AFM phase of NdSb are well reproduced by theoretical calculations, in which the Fermi-arc-like structures and sharp extra bands are originated from the in-gap surface states. However, they are demonstrated to be topological trivial. By tuning the chemical potential, the AFM phase of NdSb would go through a topological phase transition, realizing a magnetic topological insulator phase. Hence, our study sheds new light on the rare earth monopnictides for searching unusual AFM structure and the potential of intrinsic magnetic topological materials.

**Introduction**

Rare earth monopnictides (RePn) have attracted extensive attention due to their emerging topological electronic structure and extremely large magnetoresistance[1,25]. The partial filling of 4f orbitals usually leads to antiferromagnetic (AFM) ground states in some of RePn's. The AFM order consists of Ising-like moments pointing to z direction and are ferromagnetically coupled in the $x - y$ plane[6,7,8,9,10]. It provides an ideal platform to study the interaction between magnetism and topology. The topological nature in most of these materials were studied in their paramagnetic (PM) phase. For example, the PM phase of ReSb (Re=La, Ce, Pr, Sm, Gd) shows topological trivial band structures without band inversion[3-5,11-13]. While band inversion happens in the PM phase of ReBi (Re=La, Ce, Pr, Sm, Gd), resulting in a topological non-trivial semimetal phase[1,2,4,5,8,14]. Recent works of NdBi and some of RePn reported unconventional Fermi-arc states in their AFM phases[15-17], striking new interests of these materials. However, the corresponding topological nature in their magnetic ground states are rarely studied[8].

Among the RePn compounds, CeSb has been proposed to host magnetic Weyl points in its ferromagnetic (FM) phase under a magnetic field[19]. In its Ising-like stacked AFM phase, CeSb undergoes the devil's staircase transition, and the magnetic reconstruction dramatically alters the band dispersion at each transition as resolved experimentally[6]. These band reconstructions can be interpreted as band folding along the corresponding AFM wave-vectors. In contrast, exotic Fermi-arc-like surface states have been observed in the AFM phase of NdBi, which *cannot* be assigned to any folded bulk bands in calculations based on the AFM structure with single wave vector (single-q), and the origin of the splitting behavior of these surface states remains unclear[15]. However, since even in cubic systems, AFM could host several symmetric wave vectors (multi-q), such as the 3q AFM structure in USb[10,20,21], one may consider multi-q AFM configurations of NdBi. Indeed, people have provided such direction in the recent work to reproduce the unconventional Fermi-arc-like states[18]. Hence, it is crucial to identify the correlation between the AFM structure and those exotic electronic states observed in RePn, which would provide deeper understanding of the interplay between magnetism and topology in RePn systems.

Here, using angle-resolved photoemission spectroscopy (ARPES) and density functional theory (DFT) calculations, we show that another RePn compound, NdSb, goes through an unusual modulation on the electronic structures across the AFM transition. In its PM phase, NdSb is found to be a topological trivial semimetal. But in its AFM phase, new exotic surface states emerge which are most possibly related to the 2q AFM structure through our calculations. However, these surface states are demonstrated to be topological trivial. We further show that a topological phase transition can be induced by tuning its chemical potential.

**Results and Discussion**
**Antiferromagnetic order induced extra bands and DFT calculations**

NdSb crystalizes in a rock-salt structure. Resistance measurement shows a sharp AFM transition at 15.3 K (Supplementary Figure 1a), and the RRR ($(R_{300K} - R_{2K})/R_{2K}$) ratio is about 131, indicative of high sample quality. Similar as the recent work of "multi-q" magnetic structure in NdBi[18], we consider various "multi-q" structures as candidates for its magnetic ground state, for example, multiple orientations of magnetic moments can generate 1q, 2q or 3q AFM structures (configurations of these magnetic structures can be found in Figs. 1a, 1b and Supplementary Figure 2). The three dimensional (3D) Brillouin zones (BZ) and associated surface BZs in the 2q or 3q AFM phase are shown in Fig. 1c, and the blue dashed rectangles indicate the corresponding boundary of surface BZs in the PM phase (the definition of the 3D and surface BZs can be found in Supplementary Figure 1b). Our calculations show that the 2q AFM state has the lowest energy among the three kinds of magnetic structures (1q, 2q, 3q). These three magnetic structures (1q, 2q, 3q) show clear differences in the calculated electronic structures. Their calculated Fermi surfaces are plotted in Figs. 1d, 1e and 1f, respectively (more details can be found in Supplementary Figure 2). Interestingly, obvious Fermi-arc-like structures can be observed on the Fermi surfaces of both the 2q and 3q AFM structures (Figs. 1e and 1f), but not on those of the 1q AFM structure (Fig. 1d). However, the Fermi surface of the 2q AFM structure shows a 2-fold symmetry in the (010) plane of the momentum space (Fig. 1f), which is different from the 3q AFM case. Thus, the 2-fold symmetry of the Fermi surface in the (010) plane is a distinct character of the 2q AFM phase.

We have successfully cleaved two different surfaces, (001) surface (abbreviated as S1) and (010) surface (abbreviated as S2) in Figs. 1g and 1h. Our ARPES measured Fermi surface of S1 shows clear Fermi-arc-like states with 4-fold rotational symmetry (Fig.1g), however, obvious 2-fold symmetry has been observed in S2 (Fig. 1h(i)) (more details can be found in Supplementary Figure 3). Since the high symmetry $\bar{X}_{PM}$ point in the PM surface BZ is identical the second $\bar{\Gamma}$ point in the 2q AFM phase (illustrated in Fig. 1c), one can see that despite of some intensity differences of the Fermi surfaces at the $\bar{\Gamma}$ and $\bar{\Gamma}(\bar{X}_{PM})$ points (Fig. 1h(ii)), they are actually the same in principle and both could be well explained by the calculated Fermi surface in Fig. 1f(ii). What's more, we considered other possible 1q AFM structures in Supplementary Materials Fig. S4. In despite of the similar Fermi-arc-like surface states, the band structures in the particular E-k window deviate from our ARPES spectrum in Fig. 2d and 2q AFM calculations in Fig. 2f. Thus, we can conclude from our data that NdSb favors a 2q AFM magnetic ground state (More details to exclude the possibilities of other magnetic structures can be found in Supplementary Figure 4).

The calculated bulk bands in the PM and 2q AFM phases of NdSb are shown in Figs. 2a and 2b, respectively. The band inversion is absent between the Sb 5p and Nd 5d bands in the PM phase. Bulk bands in the 2q AFM phase are much more complicated due to band folding. The band structures of the PM and AFM states are obviously different in our ARPES data (Figs. 2c and 2d), and especially several sharp extra bands appear in the AFM phase. Usually, the bands induced by AFM band-folding effects have weak photoemission signal. Thus, these sharp extra bands are unlikely due to the AFM band folding. By comparing the measured (Fig. 2c) and calculated (Fig.2e) band structures of the PM phase, we find at least three pairs of sharp extra bands near the Fermi level ($E_F$) in the AFM phase (Fig. 2d). We assign these three pairs of sharp bands located around $\bar{\Gamma}(\bar{X}_{PM})$ and $\bar{\Gamma}$ points as 1, 2, 3 and 1′, 2′, 3′, respectively. One could see that they are symmetrized according to the $\bar{X}$ point in the AFM surface BZ. Pair 3/3′ forms the Fermi-arc-like states as indicated in Fig. 1h, which merge into the bulk band. This behavior is similar to the Fermi-arc-like states reported in a recent study on NdBi[15] and some other RePn[16]. Pairs 1/1′ and 2/2′ contribute to the Fermi pockets centered around $\bar{\Gamma}$. These are reproduced by DFT calculations with the 2q AFM structure (Fig. 2f), further indicating these three pairs of extra bands are most likely related to the AFM transition. The photon energy dependent measurements are presented in Supplementary Figure 5, where the bands show no dispersion along the out-of-plane momentum direction, indicative of their surface origin. In addition, there are other subtle features in the AFM phase, as indicated by the highlighted green and black dashed curves (black arrows) in Figs. 2d and 2f, including the characteristic folded bulk bands and surface states in the AFM phase at $\bar{\Gamma}$ near $E_F$, and another sharp surface state located at 0.3 eV below $E_F$ at the $\bar{\Gamma}$ point. All these emerging bands in the AFM phase are perfectly reproduced by our DFT calculations, indicating that the 2q AFM structure is possible crucial for the exotic Fermi-arc-like state and extra sharp bands. These features are unlikely from the surface effect because the surface magnetic structure is usually similar to the bulk in other RePn materials.

**Temperature evolution of the splitting behavior of the surface states**

Of particular interest is the temperature dependent behavior of the surface states across the AFM transition. Figures 3a and 3b plot the temperature evolution of the ARPES spectra at the $\bar{\Gamma}$ and $\bar{\Gamma}(\bar{X}_{PM})$, respectively. Interestingly, the two "splitting" branches of the surface states, referred as 2, 3 or 2′, 3′ in Fig. 2, gradually move towards each other and finally annihilate in the PM phase with increasing temperature. Meanwhile, pair 1/1′ shows no obvious change with increasing temperature, before it disappears in the PM phase. It should be noted that pairs 2/2′ and 3/3′ do not split with a constant energy scale along their dispersions, similar to the NdBi case[15]. This splitting behavior has also been reported in an independent work[16], we further consider the relationship between the splitting energy and the net magnetic moment of Nd. Figure 3c displays the energy distribution curves (EDC's) at various temperatures along the white dashed line in Fig. 3b, where the energy splitting between the two bands is the largest. The red arrows guide the peak splitting energy which gradually decreases with increasing temperature, and the peak separation is plotted in Fig. 3d as a function of temperature. Meanwhile, the net magnetic moment of Nd measured by previous neutron scattering measurements is also plotted in Fig. 3d for comparison[22]. The almost identical temperature dependent behaviors suggest that the splitting of the surface states is directly related to the magnetization.

**Band topology in both PM and AFM phases**

We then examine the topological nature of NdSb. A key criterion of the topological non-trivial band structure in the PM ReSb/Bi compounds is the bulk *p-d* band inversion combined with clear topological surface states at the $\bar{X}$ point[3,4,23]. In the PM phase of NdSb, the photoemission spectrum (Fig. 4a) and the corresponding EDC's (Fig. 4b) clearly show the absence of the surface states and an obvious band gap at the $\bar{X}$ point, revealing its trivial topology. Once entering the AFM phase, the bulk band structure is reconstructed due to the band folding. Interestingly, the Nd 5*d* orbitals and Sb 5*p* orbitals fold to each other resulting in series of band crossings along both the $\Gamma - X_{PM}$ direction and the $\Gamma - Z$ direction as shown in the bulk band calculations (Fig. 4c). Under different symmetries, the crossing points behave differently. As shown in the inset that enlarges the region near the crossing point along the $\Gamma - Z$ direction, two bulk Dirac points are formed characterized by different irreducible representations (LD6 and LD7). Whereas along the $\Gamma - X$ direction, the two bands (DT5) have same irreducible representations and thus a large hybridization gap (~ 40 meV) is induced by strong SOC effect (Fig. 4c). The exotic surface states are exactly located inside this band gap and merge into the bulk states.

In order to identify the topological nature in the AFM phase of NdSb, we need to calculate its topological index. We employ the method of magnetic topological quantum chemistry (MTQC)[24-27] to reveal the nature of the topology for occupied bands of NdSb, in which the number of the occupied bands is set to 76. The magnetic space group of the 2q AFM structure is Pc4$_2$/nnm or N.134.481, the corresponding indicators are defined using parity-based eigenvalues: $Z_{2p'} = 1/2Z_{4p} = \sum_{k \in TRIM} 1/4(N_k^- - N_k^+) \bmod 2$[28,29], where $N_k^-$ and $N_k^+$ are the number of odd and even parity of occupied bands at the time-reversal-invariant momenta (TRIM), respectively. The calculated eigenvalues at the eight TRIM are listed in Fig. 4d. Both $Z_{2p'}$ and $Z_{4p}$ are 0, indicating trivial topology of NdSb in its AFM phase with 76 occupied band number. Therefore, the AFM phase of NdSb is topological trivial, and the exotic in-gap surface states cannot be from a topological origin. Our findings show that they are most likely trivial surface states originated from the 2q AFM structure. Since the net magnetic moment of Nd is temperature dependent, the hybridization band gap varies with temperature, which will modify the dispersion of the in-gap surface states. This may account for the observed intriguing temperature dependent behavior of the surface states.

More interestingly, by increasing the occupied band number from 76 to 80, i.e., shift the chemical potential ~ 0.1 eV above $E_F$, a topological nontrivial phase will appear. In this case, indices $(Z_{4p}, Z_{2p'})$ are (2, 1), indicative of a strong antiferromagnetic topological insulator (Fig. 4c)[24]. This suggests that further engineering the electronic states by external effects, such as electron doping or gating, could induce a topological phase transition in the AFM structure of NdSb.

In conclusion, by performing ARPES measurements and DFT calculations, we systematically studied the electronic structures in both the PM state and the AFM state of NdSb. Exotic electronic states have been observed when it enters the AFM phase. We found that this exotic behavior is most likely related to an unconventional 2q magnetic structure. However, both its PM and AFM phase are proved to be topological trivial. Moreover, by tuning the band occupation, the AFM phase of NdSb would go through a topological phase transition to an intrinsic magnetic topological material[24,30-42]. Our results unveil the consequence of the unconventional multi-q magnetic structure on the band structure, and the interplay between band topology and magnetism in rare earth monopnictides. However, the micro-mechanism of the existence of multi-q magnetic structure in NdSb and some other rare earth monopnictides is still not clear, further experimental and theoretical researches are needed to resolve this issue.

**Note added**: While we were finalizing this manuscript, we noticed one independent study of NdSb reporting the complex band structure in the AFM phase[17] and two other independent studies of NdBi and NdSb based on the multi-q antiferromagnetic structure[16,18]. Our work considered more possibilities of magnetic structures and developed a reasonable method to address the topology of both the PM and AFM phases of NdSb.

**Methods**

**Single crystal synthesis**

Single crystals of NdSb were synthesized using indium flux method with a molar ratio of Nd: Sb: In of 1: 1: 10. The starting materials were weighted and loaded in alumina crucibles, sealed in an evacuated quartz tube, and heated to 1100 ℃ before cooled down to 800 ℃. Finally, the samples were separated from the indium in a centrifuge. The typical crystal size is $4 \times 4 \times 4$ mm.

**ARPES measurement**

ARPES measurements were performed at beamline BL13U of National Synchrotron Radiation Laboratory (NSRL) in Hefei, China. The measurement pressure was kept below $8 \times 10^{-11}$ Torr, and data were recorded by Scienta DA30 analyser at various sample temperatures. The total convolved energy and angle resolutions were 15 meV and 0.2°. The fresh surface for ARPES measurement was obtained by cleaving the NdSb sample in-situ along its natural cleavage plane.

**Computational Methods**

Electronic structure calculations were performed using density functional theory (DFT) with a plane wave basis projected augmented wave, as implemented in the Vienna *ab-initio* simulation package (VASP)[43]. The Perdew-Burke-Ernzerhof (PBE) approximation was used as the exchange-correlation potential. The 4$f$ electrons were treated as localized core electrons and spin orbit coupling was included. An energy cutoff of 300 eV and $8 \times 8 \times 8$ Γ-centered k-mesh were employed in the calculation. The surface spectra and Fermi surfaces were calculated by surface Green's function methods as implemented in WannierTools[44]. U = 7.2 eV and J = 0.7e V have been used for our DFT+U+SOC calculations in all of the considered AFM phases, resulting in a total magnetic moment of 2.7 $\mu_B$ on Nd, which is consistent with the experimental results (2.9 $\pm$ 0.2 $\mu_B$)[9]. The used U and J values were based on the localized behavior of the 4f electron of NdSb, producing similar 4f levels as the previous XPS experiments[45]. The parameters U and J modified little of the band structure near Fermi level in a relatively large range. Fermi levels were shifted up by 80 meV to match the ARPES spectra. The parity analysis is based on the method described in ref.[46].

Since the PBE is known to exaggerate the band inversion features[47], especially in ReSb/Bi compounds where the partial filled f electrons make the accurate calculations more difficult. This makes the results of PBE calculations are usually inconsistent with ARPES results[3,12]. In order to remedy this discrepancy, we slightly enlarged the lattice constant by a ratio of 1.06 to reduce the overestimated band inversion effect, which produce much better consistency of band structures between calculations and ARPES measurements. In fact, ARPES measurements have indicated the full gap features at the $X_{PM}$ point in LaSb, CeSb, PrSb, SmSb, etc[4,12]. The former theoretical calculations have noted the larger gapped feature at the $X_{PM}$ point from PrSb to YbSb due to Lathanum contraction effect[11,48], indicating a full gap must exists in the PM phase of NdSb. Therefore, our slightly lattice enlargement PBE method, providing much better consistency with ARPES, is reasonable and a potential easier method to study the topological natures in ReSb/Bi compounds.

**Data Availability**
All data needed to evaluate the conclusion in the paper are present in the paper and/or the Supplementary information.

**Acknowledgements**
This work was supported by the National Natural Science Foundation of China (Grants No. 12174362, No. 92065202, and No.11888101) and the Fundamental Research Funds for the Central Universities (Grant No. WK9990000103).

**Competing Interests**
The authors declare no competing interest.

**Author Contributions**
P. L. and T. R. L. contribute equally to this work. P. L. and J. J. conceived the experiments. P. L., R. X., Y. Z. W., J. H. Y. and J.J. carried out ARPES measurements with the assistance of S. T. C. and Z. S., T. R. L., Y. L. W. and X. G. W. performed the DFT calculations, S. L. and P. L. synthesized and characterized bulk single crystals. P. L., J. J. and D. L. F. wrote the manuscript. All authors contributed to the scientific planning and discussions.

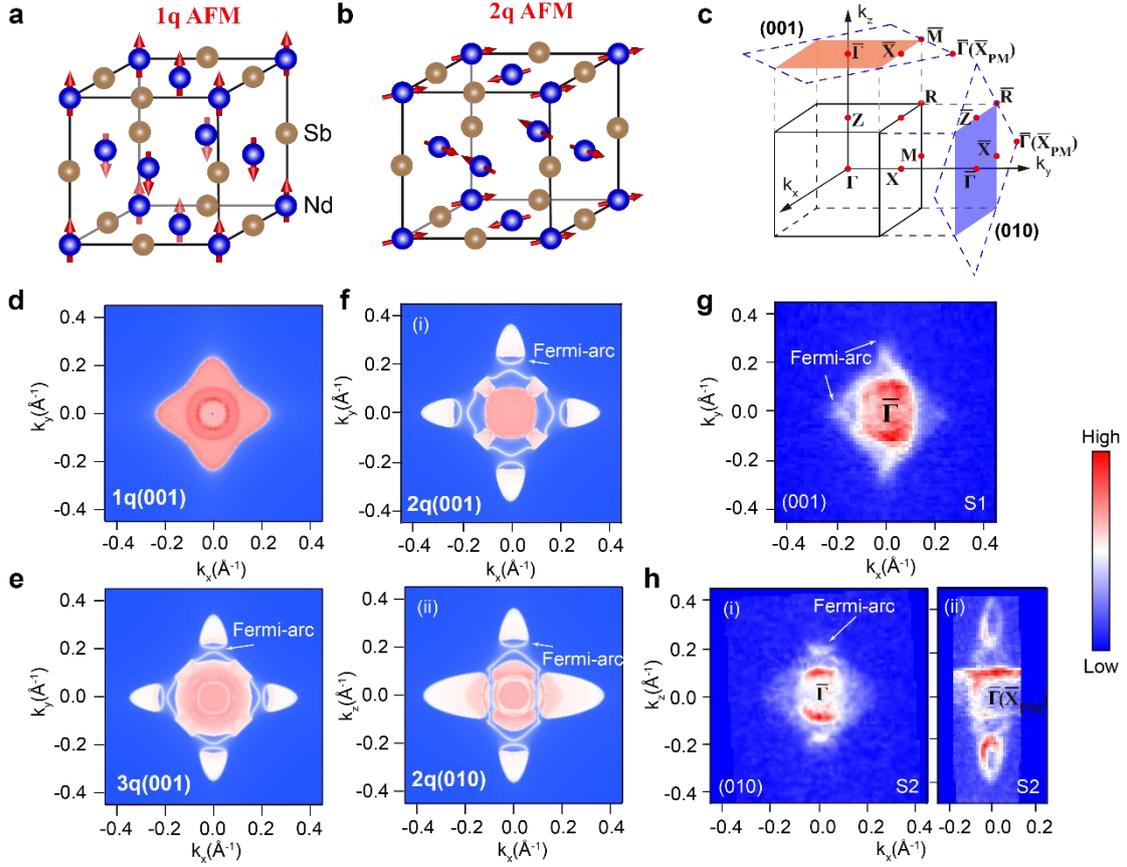

Fig. 1 **Determination of the magnetic structure of NdSb from the electronic structure.** **a** Normal 1q magnetic structure, which is similar to the cases of many other ReSb/ Bi compounds. **b** 2q magnetic structure. **c** Bulk BZ of the 2q AFM, and the projected surface BZs of the (001) and (010) surfaces with orange and light purple highlighted, respectively. The blue dashed quadrilaterals outline the surface BZ in the PM phase. **d** Calculated Fermi surface of the 1q AFM structure on the (001) surface. **e** Calculated Fermi surface of the 3q AFM structure on the (001) surface. **f** Calculated Fermi surface of the 2q AFM structure on the (001) surface (i) and (010) surface (ii). **g** ARPES measured Fermi surface of the (001) surface of S1. **h** ARPES measured Fermi surface on the (010) surface of S2 around the first $\bar{\Gamma}$ (i) and the second $\bar{\Gamma}(\bar{X}_{PM})$ (ii) in the BZ. Photon energy of 23 eV was used in ARPES measurements. The color bar shows the ARPES spectra intensity.

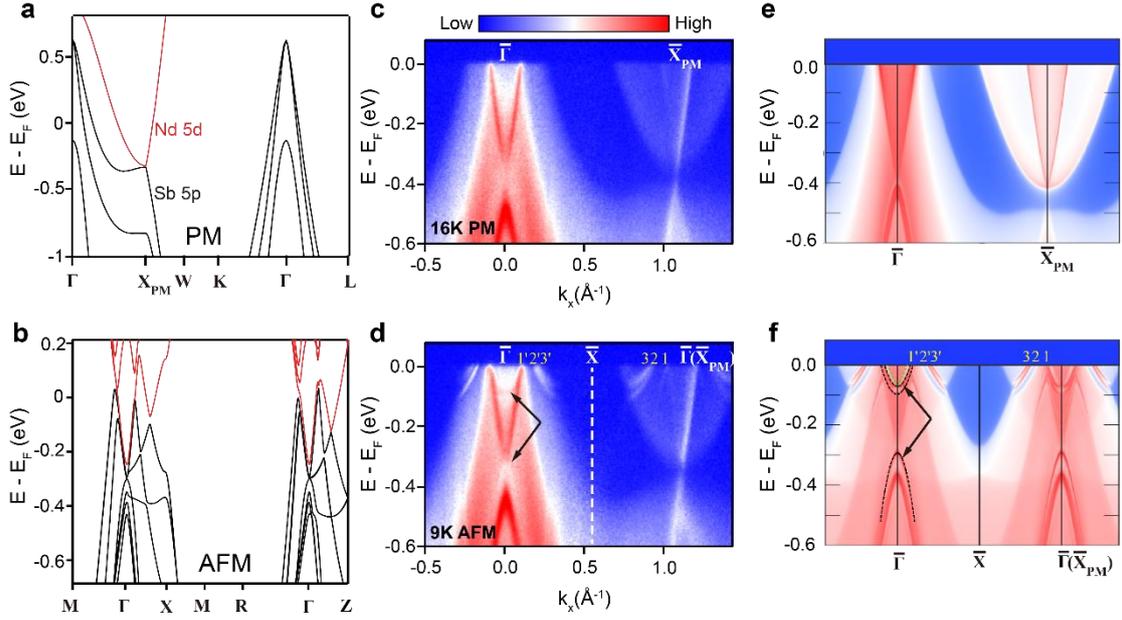

Fig. 2 **Comparison of electronic structures in the PM and AFM phase. a**, **b** DFT-calculated bulk electronic structure of NdSb in the PM phase and the 2q AFM phase, respectively. **c**, **d** ARPES spectra along $\bar{\Gamma}$ - $\bar{X}_{PM}$ direction measured at 16 K in the PM phase and 9 K in the AFM phase, respectively. Three pairs of clear surface states which are symmetrized with the $\bar{X}$ point and marked as 1, 2, 3 near $\bar{\Gamma}(\bar{X}_{PM})$ and 1′, 2′, 3′ near the $\bar{\Gamma}$ in **d**, respectively. **e**, **f** The corresponding calculations of **c** and **d**. The green and black dashed lines indicate the characteristic folded bulk bands and sharp surface states, respectively. Photon energy of 23 eV was used in the ARPES measurements. The color bar shows the ARPES spectra intensity.

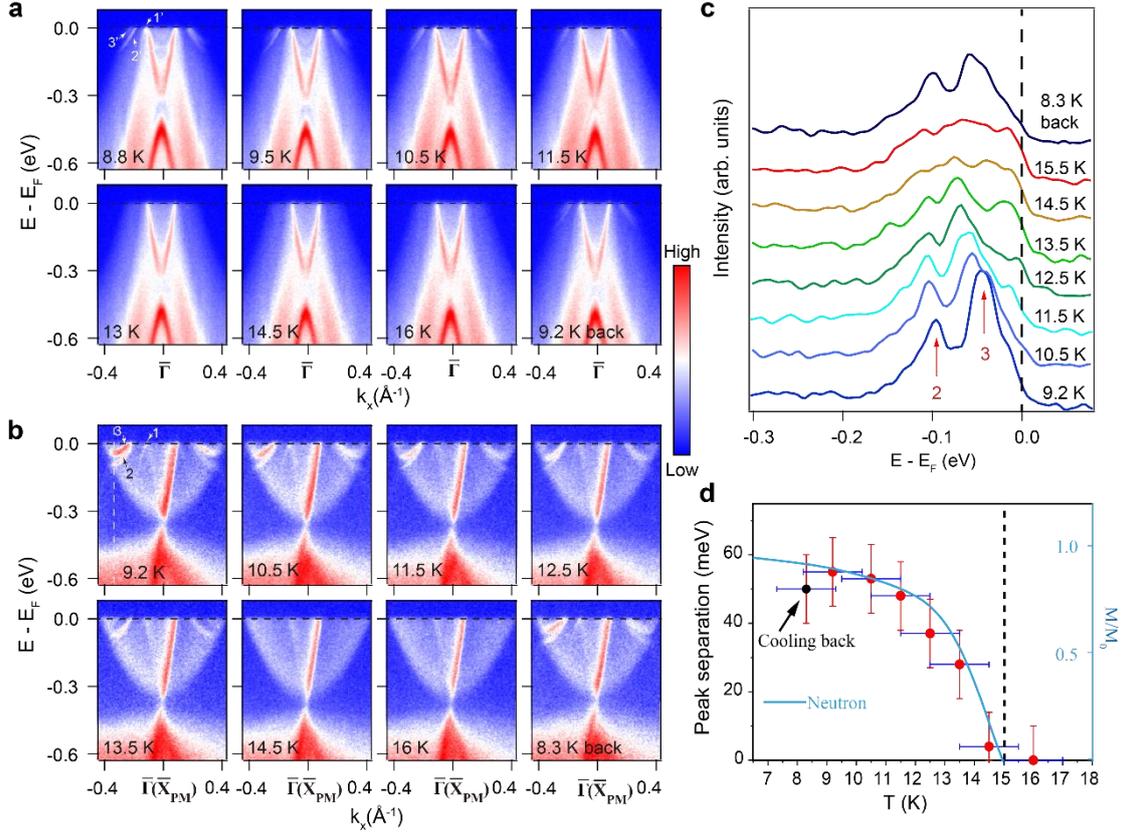

Fig. 3 **Temperature evolution of the emerging surface states. a**, **b** The temperature evolution of the spectra at the $\bar{\Gamma}$ and $\bar{\Gamma}(\bar{X}_{PM})$, respectively. The cooling-back spectrum show no obvious decay during 1 day, indicating the robustness of these surface states. **c** The extracted EDCs at different temperatures are plotted with constant intensity shift. The red arrows indicate the peaks of the splitting surface states. **d** Statistical peak separations from **c** with error bars of energy resolution (~ 10 meV) and temperature uncertainty (~1 K) are compared with the magnetic moment results from neutron measurements[22] (Reprinted from Sablik, M. J. & Wang, Y. L. *J. Appl. Phys.* **57**, 3758 (1985), with permission of AIP publishing). The color bar shows the ARPES spectra intensity.

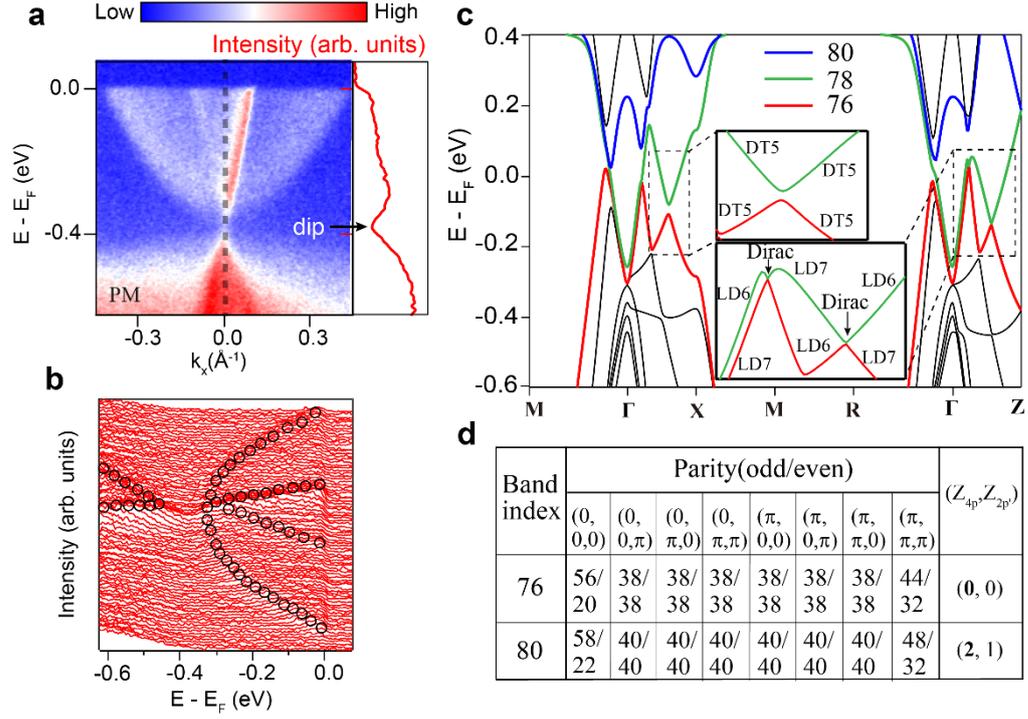

Fig. 4 **The topological identifications of the PM and 2q AFM phases of NdSb. a** ARPES spectra near the X point in the PM phase measured at 16 K with photon energy of 23 eV. The extracted EDC curve along the grey dashed line is plotted on the right side, a clear band gap is resolved. **b** The corresponding EDCs of **a** are overlapped with black circles to guide the band edge. **c** The calculated bulk bands near Fermi level are highlighted with different band index. The insets are the zoom-in view of the black dashed rectangular regions, showing two Dirac cones exist in the Γ − Z direction whereas a hybridization gap in the Γ − X direction. **d** The table lists the details of the topological indices of different band index. The color bar shows the ARPES spectra intensity.